# The two parameters (κ, r) in the generalized statistics

Guo Lina, Du Jiulin

*Department of Physics, School of Science, Tianjin University, Tianjin 300072, China*

**Abstract**

Based on the generalized Boltzmann equation and the reverse function of the distribution function, we investigate the two-parameter generalized statistics and get an expression between the two parameters ($\kappa, r$) and the physical quantities about the system considered. We find that the two parameters can define some characteristics of the system. As examples, this result can just return to the previous one obtained for Tsallis and $\kappa$ statistics. For some complex systems, we may need the two-parameter statistics to describe.

**1. Introduction**

There are often a variety of complex systems that obey asymptotic power-law distributions, such as self-gravitating systems, plasma systems, biophysical systems, economic systems and so on. In order to explain the statistical natures of such complex systems, one of the fundamental approaches is to generalize the statistical descriptions by using suitable generalizations of the Boltzmann–Gibbs–Shannon (BGS) entropy,

$$S_{BGS} = -\sum_{i=1}^{N} p_i \ln p_i . \tag{1}$$

Among the generalizations, we can find the entropies such as q-entropy [1, 2], $\kappa$-entropy [3] and the others. In 1975, Sharma and Taneja [4] and Mittal [5] derived a two-parameter entropy by generalizing Chaundy and McLeod's functional equation, which has the characteristics of Shannon' entropy. Recently, Kaniadakis et al presented two-parameter ($\kappa, r$) deformations of entropy [6], also as a generalization of Eq.(1), defined by

$$S = -\sum_{i=1}^{N} p_i \ln_{\{\kappa,r\}}(p_i) = -\sum_{i=1}^{N} p_i^{1+r} \frac{p_i^{\kappa} - p_i^{-\kappa}}{2\kappa} \tag{2}$$

From this generalized entropy, they get the two-parameter distribution function

$$f_{\{\kappa,r\}} = A \exp_{\{\kappa,r\}}\left(-\frac{mv^2}{2k_BT}\right) \tag{3}$$

where $A$ is the normalization coefficient, $k_B$ is Boltzman constant, $T$ is the temperature, and $\exp_{\{\kappa,r\}}(x)$ is the inverse function of $\ln_{\{\kappa,r\}}(x) = x^r \frac{x^\kappa - x^{-k}}{2\kappa}$. And the restriction of the two-parameter ($\kappa$, $r$) is

$$\begin{cases} -|\kappa| \leq r \leq |\kappa| & \text{if } 0 \leq |\kappa| < \frac{1}{2} \\ |\kappa| - 1 < r < 1 - |\kappa| & \text{if } \frac{1}{2} \leq |\kappa| < 1 \end{cases} \tag{4}$$

If $r = -\kappa$ and $q = 1 + 2\kappa$, (3) will become Tsallis q-distribution function [1],

$$f_q = A_q\left[1 - (1-q)\frac{mv^2}{2k_BT}\right]^{\frac{1}{1-q}} . \tag{5}$$

And if $r = 0$, (3) will become to $\kappa$-distribution function [3],

$$f_\kappa = A_\kappa\left(\sqrt{1 + \kappa^2\left(\frac{mv^2}{2k_BT}\right)^2} - \kappa\frac{mv^2}{2k_BT}\right)^{\frac{1}{\kappa}} \tag{6}$$

Some properties about the two-parameter ($\kappa, r$) entropy were investigated recently [7]. In this paper, following lines of the approach in previous works [8-10], we intend to study the physical property of the two parameters ($\kappa, r$) and their relation between the parameters and the physical quantities.

**2. Two parameters in the generalized statistics**

In order to study the relation between the two parameter ($\kappa, r$) and the physical quantities, we consider two-parameter generalized kinetic theory for the system of N particles interacting under the action of an external force field **F**. The mass of each particle is $m$. The two-parameter distribution function at the point (**x**, **v**, $t$) may be generally represented by $f_{\{\kappa,r\}}(\mathbf{x},\mathbf{v},t)$. Then the dynamical behavior of the system can be governed by the generalized Boltzmann equation,

$$\frac{\partial f_{\{\kappa,r\}}}{\partial t} + \mathbf{v}\cdot\frac{\partial f_{\{\kappa,r\}}}{\partial \mathbf{x}} + \frac{\mathbf{F}}{m}\cdot\frac{\partial f_{\{\kappa,r\}}}{\partial \mathbf{v}} = C(f_{\{\kappa,r\}}) \tag{7}$$

where $C(f_{\{\kappa,r\}})$ is the generalized collision term. According to the generalized H-theorem, any distribution function considered as the stationary-state solution of Eq.(7) must satisfy the following equation [8-10][11],

$$\mathbf{v}\cdot\nabla f_{\{\kappa,r\}} + \frac{\mathbf{F}}{m}\cdot\nabla_{\mathbf{v}} f_{\{\kappa,r\}} = 0 \qquad (8)$$

where we have used $\nabla = \frac{\partial}{\partial \mathbf{x}}$ and $\nabla_v = \frac{\partial}{\partial \mathbf{v}}$ for the convenience. When considering the two-parameter distribution function (3), we have its reverse function,

$$-\frac{mv^2}{2k_B T} = \ln_{\{\kappa,r\}}\left(\frac{f_{\{\kappa,r\}}}{A}\right) = \left(\frac{f_{\{\kappa,r\}}}{A}\right)^r \frac{\left(\frac{f_{\{\kappa,r\}}}{A}\right)^{\kappa} - \left(\frac{f_{\{\kappa,r\}}}{A}\right)^{-\kappa}}{2\kappa} = \frac{1}{2\kappa}\left(\frac{f_{\{\kappa,r\}}}{A}\right)^{r+\kappa}\left[1 - \left(\frac{f_{\{\kappa,r\}}}{A}\right)^{-2\kappa}\right]$$

(9)

so that

$$\frac{mv^2}{2k_B T}\frac{\nabla T}{T}\cdot 2\kappa = \nabla\left\{\left(\frac{f_{\{\kappa,r\}}}{A}\right)^{r+\kappa}\left[1 - \left(\frac{f_{\{\kappa,r\}}}{A}\right)^{-2\kappa}\right]\right\}$$

$$= \left[\nabla\left(\frac{f_{\{\kappa,r\}}}{A}\right)^{r+\kappa}\right]\left[1 - \left(\frac{f_{\{\kappa,r\}}}{A}\right)^{-2\kappa}\right] + \left(\frac{f_{\{\kappa,r\}}}{A}\right)^{r+\kappa}\nabla\left[1 - \left(\frac{f_{\{\kappa,r\}}}{A}\right)^{-2\kappa}\right]$$

$$-\frac{mv}{k_B T}\cdot 2\kappa = \nabla_v\left\{\left(\frac{f_{\{\kappa,r\}}}{A}\right)^{r+\kappa}\left[1 - \left(\frac{f_{\{\kappa,r\}}}{A}\right)^{-2\kappa}\right]\right\}$$

$$= \left[\nabla_v\left(\frac{f_{\{\kappa,r\}}}{A}\right)^{r+\kappa}\right]\left[1 - \left(\frac{f_{\{\kappa,r\}}}{A}\right)^{-2\kappa}\right] + \left(\frac{f_{\{\kappa,r\}}}{A}\right)^{r+\kappa}\nabla_v\left[1 - \left(\frac{f_{\{\kappa,r\}}}{A}\right)^{-2\kappa}\right]$$

(10)

Since

$$\nabla\left(\frac{f_{\{\kappa,r\}}}{A}\right)^{r+\kappa} = (r+\kappa)\left(\frac{f_{\{\kappa,r\}}}{A}\right)^{r+\kappa-1}\left[\frac{1}{A}\nabla f_{\{\kappa,r\}} - \frac{f_{\{\kappa,r\}}}{A^2}\nabla A\right] \qquad (11)$$

$$\nabla_v\left(\frac{f_{\{\kappa,r\}}}{A}\right)^{r+\kappa} = (r+\kappa)\left(\frac{f_{\{\kappa,r\}}}{A}\right)^{r+\kappa-1}\frac{1}{A}\nabla_v f_{\{\kappa,r\}}, \qquad (12)$$

substituting Eq. (11) and Eq.(12) into Eq. (8) we obtain

$$\mathbf{v}\cdot\nabla\left(\frac{f_{\{\kappa,r\}}}{A}\right)^{r+\kappa} + \frac{\mathbf{F}}{m}\cdot\nabla_v\left(\frac{f_{\{\kappa,r\}}}{A}\right)^{r+\kappa} + (r+\kappa)\left(\frac{f_{\{\kappa,r\}}}{A}\right)^{r+\kappa}\frac{1}{A}\mathbf{v}\cdot\nabla A = 0 \qquad (13)$$

And

$$\mathbf{v}\cdot\nabla\left(\frac{f_{\{\kappa,r\}}}{A}\right)^{-2\kappa}+\frac{\mathbf{F}}{m}\cdot\nabla_v\left(\frac{f_{\{\kappa,r\}}}{A}\right)^{-2\kappa}-2\kappa\left(\frac{f_{\{\kappa,r\}}}{A}\right)^{-2\kappa}\frac{1}{A}\mathbf{v}\cdot\nabla A=0 \quad .\tag{14}$$

Combining Eq.(10), Eq.(13) and Eq.(14), we can derive

$$\mathbf{v}\cdot\frac{mv^2}{2k_BT}\frac{\nabla T}{T}+\frac{\mathbf{F}}{m}\cdot\left(-\frac{m\mathbf{v}}{k_BT}\right)+(r-\kappa)\left(-\frac{mv^2}{2k_BT}\right)\mathbf{v}\cdot\frac{\nabla A}{A}+\left(\frac{f_{\{\kappa,r\}}}{A}\right)^{r+\kappa}\mathbf{v}\cdot\frac{\nabla A}{A}=0 \tag{15}$$

So we find

$$\frac{mv^2}{2k_BT}\frac{\nabla T}{T}+\mathbf{F}\left(-\frac{1}{k_BT}\right)+(r-\kappa)\left(-\frac{mv^2}{2k_BT}\right)\frac{\nabla A}{A}+\left(\frac{f_{\{\kappa,r\}}}{A}\right)^{r+\kappa}\frac{\nabla A}{A}=0 \tag{16}$$

This equation gives the relation between the two parameters ($\kappa, r$) and the physical quantities including the temperature gradient $\frac{\nabla T}{T}$, the external force $\mathbf{F}$ and the gradient normalization coefficient $\frac{\nabla A}{A}$ etc. It represents one new station-state equilibrium about the systems that can be described by two-parameter distribution function (3) between the three vectors $\frac{\nabla T}{T}$, $\mathbf{F}$ and $\frac{\nabla A}{A}$ because they must be in the same plane and satisfy Eq.(16) with the two parameters ($\kappa, r$), in some degree, it gives the physical property of two parameters ($\kappa, r$).

The physical properties of the two parameters ($r, \kappa$) are defined by the three vectors and the concrete form of the distribution function $f$ in Eq.(16), appearing to be dependent on the distribution function $f$. However, Eq.(16) is an identity for any powers of the velocity, while $f$ is defined by the parameters ($r, \kappa$), so, in principle, Eq.(16) can determine a relation between ($r, \kappa$) and the above three vectors, and it actually works for any trace form entropy. We may focus on another two-parameter ($\varsigma, \kappa$) entropy [12,13] given by

$$S_{\kappa,\varsigma}=-\sum_{i=1}^{N}p_i\ln_{\{\kappa,\varsigma\}}(p_i), \tag{17}$$

with

$$\ln_{\{\kappa,\varsigma\}}(x)=\frac{\varsigma^\kappa x^\kappa-\varsigma^{-\kappa}x^{-\kappa}-\varsigma^\kappa+\varsigma^{-\kappa}}{\kappa\varsigma^\kappa+\kappa\varsigma^{-\kappa}}. \tag{18}$$

The $q$-entropy ($S_q$, $S_{2-q}$ with $q=1+\kappa$) and the $\kappa$-entropy can be obtained as follows, $S_q=\lim_{\varsigma\to 0+}S_{\kappa,\varsigma}$, $S_{2-q}=\lim_{\varsigma\to+\infty}S_{\kappa,\varsigma}$, $S_\kappa=S_{\kappa,1}$. In this case, Eq.(16) does not depend on the distribution function and will produce the explicit formula for the parameters $q$ and $\kappa$.

## 3. Compare with Tsallis q-statistics

If take $r = -\kappa$ and $q = 1 + 2\kappa$, Eq. (3) becomes Tsallis q-distribution function,

$$f_q = A_q \left[ 1 - (1-q) \frac{mv^2}{2k_B T} \right]^{\frac{1}{1-q}}, \qquad (19)$$

and Eq.(16) becomes

$$\frac{mv^2}{2k_B T} \frac{\nabla T}{T} + \mathbf{F}\left(-\frac{1}{k_B T}\right) + (1-q)\left(-\frac{mv^2}{2k_B T}\right) \frac{\nabla A_q}{A_q} + \frac{\nabla A_q}{A_q} = 0 \qquad (20)$$

In this equation, because $\mathbf{x}$ and $\mathbf{v}$ are independent variables and Eq. (20) is identically null for any arbitrary $\mathbf{v}$, the coefficients of powers of $\mathbf{v}$ in Eq. (20) must be zero respectively. Thus, when we consider the coefficient equation for the zeroth-power terms of $\mathbf{v}$ in Eq. (20), we obtain

$$\mathbf{F}\left(-\frac{1}{k_B T}\right) + \frac{\nabla A_q}{A_q} = 0 \qquad (21)$$

For the coefficient equation of the second power of $\mathbf{v}$, we have

$$\frac{m}{2k_B T} \frac{\nabla T}{T} + (1-q)\left(-\frac{m}{2k_B T}\right) \frac{\nabla A_q}{A_q} = 0 \qquad (22)$$

Combinding Eq. (21) and Eq. (22), we find

$$\frac{\nabla T}{T} - (1-q)\mathbf{F}\frac{1}{k_B T} = 0 \qquad (23)$$

From Eqs.(21)(22)(23), we find that in Tsallis statistics, three vectors $\frac{\nabla T}{T}$, $\mathbf{F}$ and $\frac{\nabla A_q}{A_q}$ is not only in the same plane, but also at the same line. In other words, when we select $r = -\kappa$ and $q = 1 + 2\kappa$ in the generalized statistics, which just becomes Tsallis statistics, the temperature gradient $\nabla T$ and the external force $\mathbf{F}$ must be at the same line. This relation was discussed in many papers. For example, for self- gravitating systems, the external force is $\mathbf{F} = -m\nabla\varphi$, where the gravitational potential should satisfy the Poisson's equation,

$$\nabla^2 \varphi = 4\pi G m n = 4\pi G \rho \qquad (24)$$

Thus, we have the relation [10],

$$\frac{\nabla T}{T} + (1-q)\frac{m\nabla \varphi}{k_B T} = 0 \tag{25}$$

For the plasma systems, the external force is Coulombian $\mathbf{F} = e\nabla\varphi$, where the Coulombian potential satisfy the Poisson's equation,

$$\nabla^2 \varphi = -4\pi e n \tag{26}$$

Therefore, the relation (23) becomes [9]

$$\frac{\nabla T}{T} - (1-q)\frac{e\nabla \varphi}{k_B T} = 0 \tag{27}$$

The above expressions describe the physical properties of $q$-parameter in the different systems with long-range interactions, thus leading to the physical explanation for Tsallis statistics to describe the nonequilibrium stationary-state, also leading to the experimental test for Tsallis statistics from the sound speeds in the helioseismological measurements [14].

## 4. Compare with $\kappa$-statistics

If we take $r = 0$, Eq. (3) will become the distribution function in $\kappa$-statistics,

$$f_\kappa = A_\kappa \left( \sqrt{1 + \kappa^2 \left(\frac{mv^2}{2k_B T}\right)^2} - \kappa \frac{mv^2}{2k_B T} \right)^{\frac{1}{\kappa}} \tag{28}$$

and Eq.(16) will become

$$\frac{mv^2}{2k_B T}\frac{\nabla T}{T} + \mathbf{F}\left(-\frac{1}{k_B T}\right) + \left(\sqrt{1 + \kappa^2 \left(\frac{mv^2}{2k_B T}\right)^2}\right)\frac{\nabla A_\kappa}{A_\kappa} = 0 \ . \tag{29}$$

That is

$$\frac{m^2 v^4}{4k_B^2 T^2}\frac{(\nabla T)^2}{T^2} + \frac{F^2}{k_B^2 T^2} - \frac{mv^2}{k_B^2 T^2}\frac{\nabla T}{T}\cdot\mathbf{F} = \frac{(\nabla A_\kappa)^2}{A_\kappa^2} + \kappa^2 \frac{m^2 v^4}{4k_B^2 T^2}\frac{(\nabla A_\kappa)^2}{A_\kappa^2} \ . \tag{30}$$

For the same reason, the coefficient of powers of $\mathbf{v}$ in Eq. (30) must be zero respectively. The coefficient equation of the zeroth-power of $\mathbf{v}$ is

$$\frac{F^2}{k_B^2 T^2} = \frac{(\nabla A_\kappa)^2}{A_\kappa^2} . \tag{31}$$

The coefficient equation of the second power of $\mathbf{v}$ is

$$\frac{\nabla T}{T} \cdot \mathbf{F} = 0. \tag{32}$$

The coefficient equation of the fourth power of $\mathbf{v}$ is

$$\frac{(\nabla T)^2}{T^2} = \kappa^2 \frac{(\nabla A_\kappa)^2}{A_\kappa^2}. \tag{33}$$

Substitute Eq. (31) into Eq. (33), we find

$$\frac{(\nabla T)^2}{T^2} = \kappa^2 \frac{F^2}{k_B^2 T^2}. \tag{34}$$

That [15,16] is

$$\frac{|\nabla T|}{T} = |\kappa| \frac{|F|}{k_B T}. \tag{35}$$

Eq. (35) was discussed in our previous work, which describes the physical properties of $\kappa$ parameter in the discussed systems [15,16].

## 5. Conclusion

There have been several forms of entropies as the generalization of Boltzmann-Gibbs entropy. The two-parameter ($\kappa, r$) generalized entropy may unify the entropic forms. In this paper, we study the physical property of the two parameters ($r$, $\kappa$) and their relation between the physical quantities of the considered system in this generalized statistics. We use the generalized Boltzmann equation and the reverse function of the two-parameter ($\kappa, r$) distribution function to derive the relation Eq.(16). We therefore find that the physical quantities, such as the temperature gradient $\frac{\nabla T}{T}$, the external force $\mathbf{F}$ and the gradient of normalization coefficient $\frac{\nabla A}{A}$, must be in the same plane. Eq.(16) can determine a relation between the two parameters ($r, \kappa$) and the physical quantities, $\frac{\nabla T}{T}$, $\mathbf{F}$ and $\frac{\nabla A}{A}$, ans so, in principle, it defined the characteristics of the system considered.

As examples, we apply the approach of Eq.(16) in the two-parameter statistics to Tsallis and $\kappa$ framework. From Eq. (16), we can get Eq. (23), Eq.(32) and Eq.(35), which are just the results obtained in the previous works [9,10] [15,16]. Namely, in Tsallis statistics, the temperature gradient, the external force must be parallel between each other. While in the $\kappa$ statistics, the temperatur gradient is vertical to the external force. For another complex systems, we may need the two-parameters ($r, \kappa$) statistics or another one.

**Acknowledgements**

We would like to thank the National Natural Science Foundation of China under grant No.10675088 for the financial support.